\documentclass{JHEP3}

\usepackage{epsfig,amssymb,amsmath,latexsym}

\newcommand{\be}{\begin{equation}}
\newcommand{\ee}{\end{equation}}
\newcommand{\bea}{\begin{eqnarray}}
\newcommand{\ena}{\end{eqnarray}}

\newcommand{\de}{\partial}

\newcommand\E{{\mathcal{E}}}

\newcommand\K{{\mathcal{K}}}

\newcommand\M{{\cal M}}
\renewcommand\P{{\cal P}}
\newcommand\C{{\cal C}}
\renewcommand\l{\lambda}
\renewcommand\L{\Lambda}
\renewcommand\o{\omega}
\renewcommand\a{\alpha}
\renewcommand\b{\beta}
\renewcommand\k{\kappa}
\newcommand\e{\epsilon}
\newcommand\m{\mu}
\newcommand\n{\nu}
\newcommand\g{\gamma}
\renewcommand\r{\rho}
\newcommand\s{\sigma}

\newcommand\tr{\text{tr}}
\newcommand\PM[1]{\begin{pmatrix}#1\end{pmatrix}}

\renewcommand\l{\ensuremath{\lambda}}

\renewcommand\L{\ensuremath{\Lambda}}
\newcommand\diag{\text{diag}}
\newcommand\ba{\begin{array}}
\newcommand\ea{\end{array}}

\title{Exact Spherically Symmetric Solutions in Massive Gravity }
\date{\today}

\author{Z. Berezhiani,\!$^a$ D. Comelli,\!$^b$ F. Nesti,\!$^a$ L. Pilo$^a$\\
  \llap{$^a$}Dipartimento di Fisica, Universit\`a di L'Aquila, I-67010 L'Aquila, and\\
  INFN, Laboratori Nazionali del Gran Sasso, I-67010 Assergi, Italy\\
  \llap{$^b$}INFN, Sezione di Ferrara,  I-35131 Ferrara, Italy}

\abstract{A phase of massive gravity free from pathologies can be
  obtained by coupling the metric to an additional spin-two field.  We
  study the gravitational field produced by a static spherically
  symmetric body, by finding the exact solution that generalizes the
  Schwarzschild metric to the case of massive gravity.  Besides the
  usual $1/r$ term, the main effects of the new spin-two field are a
  shift of the total mass of the body and the presence of a new
  power-like term, with sizes determined by the mass and the shape
  (the radius) of the source.  These modifications, being source
  dependent, give rise to a dynamical violation of the Strong
  Equivalence Principle.  Depending on the details of the coupling of
  the new field, the power-like term may dominate at large distances
  or even in the ultraviolet. The effect persists also when the
  dynamics of the extra field is decoupled.}

\begin{document}

\section{Introduction}

The search for large-distance modified theories of gravity,
motivated by the evidence for the cosmological acceleration, has
stimulated a number of studies in the recent years.  The main goal has
been to look for a massive deformation of standard general relativity,
featuring a large distance (infrared) modification of the Newtonian
gravitational potential, and massive gravitons.

The idea of considering a Lorentz-invariant theory of a massive
spin-two field dates back to 1939~\cite{PF}: the resulting theory is
plagued by a number of diseases that make it unphysical, besides being
phenomenologically excluded. In particular, the modification of the
Newtonian potentials is not continuous when the mass $m^2$ vanishes,
giving a large correction (25\%) to the light deflection from the sun
that is experimentally excluded~\cite{DIS}. A possible way to
circumvent the physical consequences of the discontinuity was proposed
in~\cite{Vainshtein}; the idea is that in the Fierz Pauli theory (FP)
the linearized approximation breaks down near the star and an improved
perturbative expansion must be used, leading to a continuous result
when $m \to 0$.  Whether the solution associated with the improved
perturbative expansion valid near the star can be extended up to
infinity is an open problem~\cite{DAM}. In addition, FP is problematic
as an effective theory at the quantum level.  Regarding FP as a gauge
theory where the gauge symmetry is broken by a explicit mass term $m$,
one would expect a cutoff $\Lambda_2 \sim m g^{-1} = (m
M_{pl})^{1/2}$, however the real cutoff is $\Lambda_5 = (m^4
M_{pl})^{1/5}$~\cite{NGS} much lower than $\Lambda_2$. A would-be
Goldstone mode is responsible for the extreme ultraviolet sensitivity
of FP theory, that becomes totally unreliable in the absence of proper
completion.  These issues cast a shadow on the the possibility of
realizing a Lorentz-invariant theory of massive gravity~\cite{VAIN}.

It was recently noted that by allowing lorentz-breaking mass terms for
the graviton the resulting theory can be physically viable~\cite{RUB},
being free from pathologies such as ghosts or low strong coupling
scales, while still leading to modified gravity.  Since the mass terms
break anyway the diffeomorphisms invariance, this possibility was
analized mainly in a model-independent way, by reintroducing the
goldstone fields of the broken gauge invariance, and by studying their
dynamics~\cite{NGS,DUB}; we refer to a recent review for the status
and results in this direction~\cite{RUB-TIN}.  This approach has the
power of being model independent, but the advantage turns into a
difficulty when investigating the concrete behavior of solutions.

In~\cite{us} we considered a class of theories that generate
Lorentz-breaking mass terms for the graviton, by coupling the metric
to an additional spin-two field.  This system was originally
introduced and analyzed by Isham, Salam and Strathdee~\cite{Isham,IS}
and reanalyzed more recently in~\cite{DAM1,rotw,peloso}.  While this
approach may seem antieconomical, we stress that this is the simplest
model that can explain dynamically the emergence of lorentz breaking
and give mass to the graviton. What happens is that the two tensor
fields lead in general to two coexisting and different backgrounds,
inducing Lorentz-breaking mass terms at linearized order. For a
general discussion on the consequences of Lorentz Breaking
see~\cite{kostelecky}.

The linearized analysis showed that only two gravitons propagate, one
massive and the other massless, both with two polarization states,
representing two kinds of gravitational waves (GW).  These are the
only states in the theory that feel the Lorentz breaking, showing a
frame-dependence that may be measured at future GW detectors.

In addition, the linearized gravitational potential differs in a
crucial way from the Newtonian one: it contains a new term that is
linearly growing with distance. Of course, this signals the breakdown
of perturbation theory at large distances, and in this regime the
theory should be treated fully nonlinearly. This fact is not
surprising, since one has effectively introduced nonlinear
interactions, and therefore antiscreening may be present also at
classical level like in non-abelian gauge theories.  We believe that
this is a general feature of massive gravity theories due to the
presence in the full theory of nonderivative interaction terms.  In
such situations the linearized analysis is of limited reach, and we
are forced to find exact classical solutions to be compared with the
standard Schwarzschild metric.  Though in general this a very hard
task, we have managed to find a whole class of interaction terms for
which nontrivial and rather interesting exact solutions can be found.

After describing the setup and the flat backgrounds in
section~\ref{sec:model} ad~\ref{sec:background}, we review the
linearized analysis and its problems in section~\ref{sec:lin}. We then
describe the spherically symmetric solutions in
section~\ref{sec:spherically}, that we match with an interior star
solution to estimate the modifications of the gravitational potential
as a function of the source parameters.  We also comment on the
properties of these solutions and of the whole theory in the
interesting limit when the second metric decouples, leaving just one
massive gravity theory with modified Schwarzschild solutions, as well
as in the Lorentz-invariant limit.

\section{The model}
\label{sec:model}

Consider a gravity theory in which, besides our standard metric field,
an additional rank-2 tensor is introduced in in the form of a bimetric
theory. The action is taken as\footnote{We use the mostly plus
  convention for the metric.  Indices of type 1(2) are raised/lowered
  with $g_1(g_2)$.}
\be
S = \int \!\!d^4 x\, \left[ \sqrt{-g_1} \left(M_{pl1}^2 R_1+ {\cal L}_{1}
  \right) + \sqrt{-g_2} \left( M_{pl2}^2 R_2+{\cal L}_{2}\right) - 4 (g_1
  g_2)^{1/4} V(X) \right]\,,
\ee
and for symmetry each rank-2 field is coupled to its own matter with
the respective Lagrangians ${\cal L}_{1,2}$. In the interaction term
we only consider non-derivative couplings.  The only invariant tensor,
without derivatives, that can be written out of the two metrics is
$X^\mu_\nu = g^{\mu \alpha}_1 {g_2}_{\alpha \nu}$, and then $V$ is
taken as a function of the four independent scalars $\{\tau_n =
\tr(X^n),\ n = 1,2,3,4 \}$ made out of $X$.  The cosmological
terms can be included in $V$, e.g.\ $V_{\Lambda_1} = \Lambda_1
q^{-1/4}$, with $q =\det X=g_2/g_1$.

Then the (modified) Einstein equations read
\begin{gather}
\label{eqm1}
M_{pl1}^2 \,{E_1}^\mu_\nu  +  Q_1{}^\m_\n = \frac{1}{2}\, {T_1}^\mu_\nu   \\
\label{eqm2}
M_{pl2}^2 \, {E_2}^\mu_\nu  +  Q_2{}^\m_\n = \frac{1}{2}\, {T_2}^\mu_\nu   \, ,
\end{gather}
where we defined the effective energy-momentum tensors induced by the
interaction:
\bea
\label{eq:q1}
 Q_1{}_\nu^\mu &=&  q^{1/4} \left[ V \delta^\mu_\nu- 4 (V'X)^\mu_\nu  \right]  \\[1ex]
\label{eq:q2}
 Q_2{}_\nu^\mu &=&  q^{-1/4} \left[ V \delta^\mu_\nu+ 4 (V'X)^\mu_\nu  \right] ,
\ena
with $(V^\prime)^\mu_\nu = \de V / \de X_\mu^\nu$.  Indeed, the field
$g_2$ plays the role of matter in the equations of motion for $g_1$,
and viceversa for $g_1$.

The Einstein tensors satisfy the corresponding contracted Bianchi
identities\footnote{$\nabla_{1/2}$ denotes the covariant derivative
  associated to the Levi-Civita connection of $g_{1/2}$.}
\be
g_1^{\alpha \nu} {\nabla_1}_\alpha {E_1}_{\mu \nu} = \nabla_1^\nu {E_1}_{\mu \nu} =0   \qquad 
g_2^{\alpha \nu} {\nabla_2}_\alpha {E_2}_{\mu \nu} = \nabla_2^\nu {E_2}_{\mu \nu} =0 \, .
\ee
that follows from the invariance of the respective Einstein-Hilbert
terms under common diffeomorphisms
\be
\delta {g_1}_{\mu \nu} = 2 {g_1}_{\a(\m}{\nabla_1}_{\n)}  \xi^\a 
\qquad 
\delta {g_2}_{\mu \nu} = 2 {g_2}_{\a(\m}{\nabla_2}_{\n)}  \xi^\a \,.
\ee
The interaction term is also separately invariant and we can derive
conservation laws for $Q_1$ and $Q_2$ similar to the conservation of
the energy-momentum tensor in GR:
\be
\begin{split}
& \nabla_1^\nu {Q_1}_{\mu \nu} =0  \qquad \text{on shell for } g_2  \\
& \nabla_2^\nu {Q_2}_{\mu \nu} =0  \qquad \text{on shell for } g_1 \, .
\end{split}
\label{cons}
\ee
These identities are quite powerful; for instance they allow to solve
completely the simplest of these models, when $V$ is a function of $q$
only.  This peculiar case is discussed in
appendix~\ref{sec:simplest}.

\subsection{Asymptotic solutions}
\label{sec:background}

At infinity, far from all the sources, we expect that $g_1$ and $g_2$
are maximally symmetric, setting up the benchmark for the asymptotic
behavior of all solutions of the EOM.  Denoting with $- \K_a /4$ the
constant scalar curvature of $g_{a\, \mu\nu} $, i.e.
\be
M_{pl\,a}^2 \, {E_a}_{\mu \nu} = \K_a \, {g_a}_{\mu\nu}  \,,\qquad (a=1,2)
\ee
the equations (\ref{eqm1})-(\ref{eqm2}) read
\begin{gather}
\label{cc1}
2 V + \left(q^{-1/4} \mathcal{K}_1 +   q^{1/4} \mathcal{K}_2 \right)=0  \\
\label{cc2}
8 \left(V^\prime X \right)^\mu_\nu + \delta^\mu_\nu \, 
\left(q^{1/4} \mathcal{K}_2- q^{-1/4} \mathcal{K}_1  \right)=0 \,. 
\end{gather}
and these equations can be solved for specific ans\"atze.  In order to
study the properties of this model for asymptotically flat spaces, we
analyze first the biflat solutions, $\mathcal{K}_1 = \mathcal{K}_2
=0$. Eqs.~(\ref{cc1})-(\ref{cc2}) yield:
\begin{gather}
\label{bc}
V^\prime{}_\m^\n=0\,, \qquad V=0\,.
\end{gather}
Assuming that rotational symmetry is preserved and that the two
metrics have the same signature, the biflat solution can written in
the following form
\be
\begin{split}
& {\bar g}_{1\,\mu\nu} = \eta_{\m\n} \equiv \diag(-1,1,1,1) \\[.5ex]
& {\bar g}_{2\,\mu\nu} = \omega^2\,\diag(-c^2,1,1,1)\,,
\label{vac}
\end{split}
\ee
where $c$ parametrizes the speed of light in sector 2 and $\omega$ is
the relative conformal factor.  Eqs.~(\ref{bc}) correspond to three
independent equations $V=0$, $V^\prime{}_0^0=0$ and
$V^\prime{}_i^i=0$, where $0$ and $i=1,2,3$ stand for temporal and
spatial indices.  Therefore, two of these equations determine the
values of two parameters $c$ and $\omega$, while the third represents
a fine tuning condition for the function $V$, necessary to ensure
flatness. The same sort of fine tuning is necessary in the context of
normal GR to set the cosmological term to zero. Therefore, for a
generic function $V(X)$ we expect to have a Lorentz Breaking (LB)
solution with $c\neq1$, and hence a preferred reference
frame~(\ref{vac}) in which both metrics are diagonal. Asymptotic flat
solutions should approach~(\ref{vac}) in a suitable coordinate system.
In addition, there exists also a Lorentz Invariant (LI) solution, with
$c=1$: in this case two equations coincide
$V^\prime{}_0^0=V^\prime{}_i^i$, and can be used to determine the
value of~$\o$.

Summarizing, asymptotically the solutions fall in two branches: LI
with $c=1$, and LB with $c\neq1$.\footnote{ In the special case when
  $V=V(\det X)$, Bianchi identities force $\det X$ to be constant, and
  there are additional gauge symmetries that allow to set $c=1$. As a
  result the branches are equivalent (appendix~\ref{sec:simplest}).}
The LB branch is of particular interest since it naturally allows for
consistent massive deformations of gravity~\cite{us}.

\subsection{Review (and critics) of the linearized analysis}
\label{sec:lin}

In~\cite{us} we performed a linearized analysis around the biflat
background, that we report here for the LB branch.

In addition to the kinetic terms, the linearized action contains a
Lorentz-breaking mass term for the fluctuations
$h_{a\,\m\n}=g_{a\,\m\n}-\bar g_{a\,\m\n}$ ($a=1,2$). Since on the
biflat background $V=V'{}_\m^\n=0$, one can expand the potential at
second order in the fluctuations $Y = \bar X {\bar g}_2^{-1} h_2 -
{\bar g}_1^{-1} h_1 \, \bar X$, and define the mass lagrangian:
\bea
{\cal L}_m&=& 
-2  \left( {\bar g}_1 {\bar g}_2 \right)^{1/4} 
 \, \tr \left[ Y\, V''(\bar X) \, Y \right] \nonumber\\
 &\equiv& \frac14\Big( h_{00}^t\, \M_0\, h_{00} + 2 h_{0i}^t\, \M_1\, h_{0i}  
     -  h_{ij}^t\, \M_2\, h_{ij} + h_{ii}^t\, \M_3\, h_{ii} 
     - 2  h_{ii}^t\, \M_4\, h_{00} \Big)\,.
\label{eq:mass}
\ena
Here $h_{\m\n}=\{h_{1\m\n},h_{2\m\n}\}$ is the column vector of
fluctuations and $\M_{0,1,2,3,4}$ are 2$\times$2 mass matrices. It is
then crucial to realize that, due to linearized gauge invariance (that
we remark is never broken) these matrices are of rank-one; one can write
\bea
 \nonumber  \M_0 &=& \l_0 \, \C^{-2} \P \C^{-2} \\
\label{eq:masses}
\qquad \M_{2,3} &=&  \l_{2,3}\,\, \P\,,\qquad \qquad
\P \equiv \PM{\ 1 &  -1 \\ -1\, & \ 1}\,,\quad   \C \equiv \PM{\ 1\  & \\  & \ c\ }  \\
\M_4 &=& \l_4\, \C^{-2} \P  
 \nonumber  
\ena
where $\l_{0,2,3,4}$ depend on the potential. In addition, due to the
LB, $\M_1$ vanishes regardless of the potential $V$.  This fact leads
to a well defined phase of linearized massive gravity.\footnote{The
  vanishing of the second eigenvalue of $\M_1$ can be understood by
  noting that $h_1{}_{0i}-h_2{}_{0i}$ is a goldstone direction,
  corresponding to the broken boosts in the LB background. In
  sec.~\ref{sec:localLB} we comment on the fate of this condition on
  nontrivial backgrounds.}

In this phase, the only propagating states are the two spin-2 tensor
components of the fluctuations (two polarizations each) corresponding
to two gravitons, of which one is massless and the other has mass
$\l_2$.  Their dispersion relation is non-linear due to their mixing
and their different propagating speeds~\cite{us}.

The other components, scalars and vectors, do not propagate, and
therefore discontinuity and strong coupling problems are absent in
this phase.  They however mediate instantaneous interactions, so the
Newtonian potentials that one finds at linearized level are then
drastically modified; for example in sector 1, the potential from a
point-like source $M_1$ is:
\be
\label{eq:muterm}
\Phi_{1}=-\frac{GM_1}{r}+GM_1\m^2r\,,
\ee
%
where 
\be
\label{eq:mu}
\m^2\equiv \frac{\l_2}{2M_1^2} \frac{3\l_4^2-\l_0(3\l_3-
\l_2)}{\l_4^2-\l_0(\l_3- \l_2)}
\ee
and for later reference note that $\m^2$ may be negative.

 The linearly growing term in~(\ref{eq:muterm}) signals the breakdown
 of perturbation theory at distances larger than $r_{IR}= (G M_1\m^2
 )^{-1}$ ~\cite{DUB,us}, and one usually considers the solution to be
 valid as long as the potential stays in the weak field
 regime. However, one should note that the linear term in
 (\ref{eq:muterm}) is induced by an other scalar field having an
 instantaneous interaction and acting as a source for $\Phi$ (see
 e.g.~\cite{DUBA,blas}). It is then easy to realize that non-linear
 corrections to this field can drastically modify the IR behavior,
 even in the weak-field regime.  We can clarify this point by showing,
 as an example, two systems of differential equations that differ by
 nonlinear terms and have drastically different IR behavior
\be
\left\{\ba{l} \Delta \Phi+\m^2\sigma=M\delta^3(x)\\\Delta \sigma=M\delta^3(x) \ea\right.
\qquad\quad
\left\{\ba{l} \Delta \Phi+\m^2\sigma=M\delta^3(x)\\\Delta\sigma+\l\,\sigma^2=M\delta^3(x)\,,
\ea\right.
\ee
Here $\Phi$ is a scalar field (mimicking the gravitational potential)
$\sigma$ is an additional scalar field coupled to it by a mass term
(and $\Delta$ is the laplacian). While in the first system $\s\sim
M/r$ and this induces a linear term in $\Phi$ like
in~(\ref{eq:muterm}), in the second system $\sigma$ drops to zero
faster than $M/r$ so that the bad behavior of $\Phi$ is cured. What
happens is that the IR behavior is dominated by a non-linear term,
because effectively $\Delta\to0$ at large distances.  We incidentally
point out that standard GR is safe in this respect, because nonlinear
terms coming from the Einstein tensor are always accompanied by two
derivatives and thus are equally suppressed at large distances.

We are thus led to the conclusion that in massive gravity the
situation is similar to non-abelian gauge theories, where the large
distance behavior is generically non-trivial and inaccessible to the
linearized approximation.  We recall that in Yang-Mills theories,
nonabelian configurations of charges can lead to non coulomb-like
classical solutions, screening or even anti-screening the charge,
leading also to infinite energy configurations~\cite{adler}.

In this situation what one may try is to really look at higher orders
and maybe retain the first terms that are relevant at large distance.
This approach would require the painful procedure of defining the
gauge invariant fields at higher orders, and would also lead to
nonlinear terms mixing scalars, vectors and tensors. Instead of
following this approach, we find more instructive to study the exact
spherically symmetric solutions.

\section{Exact spherically symmetric static solutions}
\label{sec:spherically}

The Schwarzschild solution describes the spherically symmetric
gravitational field produced by a spherically symmetric source.  It is
crucial to understand what kind of modification is introduced in this
theory by the presence of a new spin 2 field.  Spherical symmetry
allows us to choose a coordinate patch $(t,r,\theta, \varphi)$ where
$g_1$ and $g_2$ have the form
\begin{gather}
\label{sm1}
ds^2_1 = - J \, dt^2 + K \, dr^2 + r^2 \, d \Omega^2  \\[.5ex]
ds^2_2 = - C \, dt^2 + A \, dr^2 + 2 D \, dt dr + \, B \, d \Omega^2 \,. 
\label{sm2}
\end{gather}
and all the functions $J, K, C, A, D, B$ entering $g_1$ and $g_2$ are
function of $r$ only. Notice that the off-diagonal piece $D$ cannot be
gauged away.

\subsection{Black hole solutions}

In the absence of matter, a number of interesting properties follow
from the form of the Einstein tensors $E_1{}_\nu^\mu$, $E_2{}_\nu^\mu$
derived from (\ref{sm1})-(\ref{sm2}) and do not depend on the chosen
$V$. Following~\cite{IS,chelaflores} the spherically symmetric
solutions can be divided in two classes: type I with $D \neq 0$ and
type II with $D = 0$. We shall focus here mainly on type I solutions.

\pagebreak[3]

Since $E_1{}_\nu^\mu$ is diagonal by the choice of the first metric,
then also $(V'X)^\mu_\nu$ must be diagonal because of the
EOM~(\ref{eqm1}). The only possible source of a off-diagonal term in
the RHS of (\ref{eqm2}) would be $(V'X)^\mu_\nu$, so as a result also
$E_2{}_\nu^\mu$ must be diagonal, i.e.\ $E_2{}_r^t=0$. For type I
solutions, this condition amounts to a single equation:
\be
\label{eq:det}
AC+D^2= d_2\frac{(B')^2}{4B}\,,
\ee
where $d_2$ is a constant. Incidentally using this relation it turns
out that ${E_2}_t^t={E_2}_r^r$, then using (\ref{eqm2}) also
$(V'X)^t_t= (V'X)^r_r$, and by (\ref{eqm1}) we have also that
${E_1}_t^t={E_1}_r^r$. This relation determines $K$ in terms of $J$
\be
K=\frac{d_1}{J}\,,
\ee
with $d_1$ an other constant.  

The metric 2 can be brought in a diagonal form by a coordinate change
$dt=dt'+dr\,D/C$.  Thanks to (\ref{eq:det}), in the new coordinates we
have
\be
\label{eq:g2schw}
ds_2^2 =- C \,  {dt^\prime}^2  + \frac{(B')^2}{4B}\frac{d_2}{C} \, dr^2 + B \, d \Omega^2 
\ee
(and of course the metric 1 in no longer diagonal). Then by a suitable
change of $r$ the metric 2 can also be put in a Schwarzschild-like
form; setting $r'=\sqrt{B(r)}$, we find
\be
\label{eq:g2schw2}
ds_2^2 =- C \,  {dt^\prime}^2  + 
\frac{d_2}{C} \, {dr^\prime}^2 + r^{\prime2} \, d \Omega^2 \, , \qquad  C(r)=C(r') \,,
\ee
which shows that $C$ is the physically relevant potential in sector 2.

\medskip

To proceed further a choice of $V$ is needed. In the existing
literature essentially all the results are based on a potential
$V_{\text{IS}}$ introduced in~\cite{zumino} and~\cite{Isham} in the
context of hadronic physics.\footnote{In the years preceding QCD, the
  proposal of Isham, Salam and Strathdee was that of a second metric
  mediating a strongly coupled interaction, responsible for
  confinement of quarks inside tiny black holes.}  The motivation for
this choice is probably due to the fact that $V_{\text{IS}}$ is the
simplest potential producing a FP mass term in the (Lorentz-invariant)
linearized limit:
\bea
V_{IS}     &=&(\tau_{-2}-\tau_{-1}^2+6\tau_{-1}-12)\nonumber\\
          &=&(g^{2\,\m\n}-g^{1\,\m\n})(g^{2\,\r\s}-g^{1\,\r\s})(g_{1\,\m\r}g_{1\,\n\s}-g_{1\,\m\n}g_{1\,\r\s})\\
          &\simeq& \tr(h_-^2)-\tr(h_-)^2\qquad\qquad \text{for}\ \bar g_1=\bar g_2=\eta\,. \nonumber
\ena
where $h_-{}_{\m\n}=h_2{}_{\m\n}-h_1{}_{\m\n}$.
For $V_{\text{IS}}$ it was shown in~\cite{IS} that type-I solutions
 are always Schwarzschild-(A)dS, and it was recently realized that
 these solutions are present for any potential~\cite{blas}.  
It turns out that $V_{\text{IS}}$ can be deformed and there exists a
whole family of potentials for which the exact spherically symmetric
solutions can be found.  

Let us consider the family of potentials
\bea
\label{eq:genpot}
V&=&  a_0+a_1 V_1+a_2V_2+a_3V_3+a_4V_4+b_1V_{-1}+b_2V_{-2}+b_3V_{-3}+b_{4}V_{-4}  \nonumber\\
 &&{}+  q^{-1/4} \Lambda_1 + q^{1/4} \Lambda_2 \,, 
\ena
where we introduced the following combinations involving the
generalized determinants (again $\tau_n = \tr(X^n)$ and
$\epsilon$ is the 4-index antisymmetric symbol)
\be
\label{eq:potentials}
\begin{split}
& V_0 = \frac{1}{24|g_2|}(\e\e g_2g_2g_2g_2)= 1 \equiv \frac{1}{24 q}(\tau_{1}^4- 6 \, \tau_{2}\tau_{1}^2+8 \, \tau_{1}\tau_{3}+3 \, \tau_{2}^2-6 \, \tau_{4} )\\
& V_1 = \frac{1}{6|g_2|}(\e\e g_2g_2g_2g_1)  =(\tau_{-1}) \equiv \frac{1}{6 q}(\tau_1^3-3 \, \tau_2\tau_1+2 \, \tau_3)\\
& V_2 = \frac{1}{2|g_2|}(\e\e g_2g_2g_1g_1)  =(\tau_{-1}^2-\tau_{-2}) \equiv q^{-1}(\tau_1^2-\tau_2)  \\
& V_3 = \frac{1}{|g_2|}(\e\e g_2g_1g_1g_1)   =(\tau_{-1}^3-3\tau_{-2}\tau_{-1}+2 \, \tau_{-3}) \equiv 6\, q^{-1} \, \tau_1 \\
& V_4 = \frac{1}{|g_2|}(\e\e g_1g_1g_1g_1)   = (\tau_{-1}^4-6 \, \tau_{-2}\tau_{-1}^2+8 \, \tau_{-1}\tau_{-3}+3 \, \tau_{-2}^2-6 \, \tau_{-4}) \equiv 24\, q^{-1} 
\\ 
\end{split}
\ee
and where $V_{-n}=V_n(X\to X^{-1})$.  The cosmological constants
$\Lambda_1$ and $\Lambda_2$ have been added to simplify the asymptotic
flatness conditions.  The Isham-Storey potential is recovered by
setting $a_0=-12$, $a_1=6$, $a_2=1$ and $b_1=b_2=b_3=b_4=a_3=a_4=0$.

Remarkably, the general combination $V$ of (\ref{eq:genpot}) leads to
solvable equations for type I spherically symmetric solutions, and
these can be found in a closed form (these equations are the main
result of the paper):\footnote{This solvability is linked to the
fact that the combinations $V_n$ are actually the coefficients of the
secular equation of $X$ and are (multi)linear combinations of its
eigenvalues $\l_i$:
$V_n=\sum_{i_1>i_2\cdots>i_n}\l_{i_1}\l_{i_2}\cdots\l_{i_n}$.}
\bea
\label{eq:exterior}
J&=&\Big[1-2\frac{Gm_1}{r}+ \K_1r^2\Big]+ 2G\,S\, r^{\gamma}\,,
\qquad\qquad \qquad\qquad\quad
KJ=1\,,\\[1ex]
C&=&c^2\omega^2 \Big[1-2\frac{Gm_2}{\kappa\,r} + \K_2 r^2 \Big] -
        \frac{2G}{c\,\o^2\kappa}S\, r^{\gamma}\,,
\qquad\quad
D^2+AC=c^2\omega^4\\[1ex]
B&=&\omega^2r^2\,,
\qquad\qquad\qquad 
A= \o^2 \frac{\tilde J-\tilde C-\tilde J\, \tilde S\, r^{\gamma-2} }{\tilde J^{2}}\,,
\ena
with $\{\tilde J,\tilde C\}=\{J,C\}/\o^4(c^2+1)$,
$\tilde S=S/\l_2\,[(c^2-1)(\g+1)(\g-2)/16\o^2c^{1/2}(c^2+1)]$.
The solution depends on the integration constants $m_1, m_2$ and $S$
and we have introduced $G= 1/16 \pi M_{pl1}^2$ and $
\k=M_{pl2}^2/M_{pl1}^2$.  The values of $c^2$, $\g$ and of the
graviton mass $\l_2/M_{pl1}^2$ of the linearized analysis
(\ref{eq:masses}) are given in terms of the coupling constants:
\be
\label{eq:gengamma}
 c^2=-\frac{\tilde a_1+4\tilde a_2+6\tilde a_3}
           {\tilde b_1+4\tilde b_2+6\tilde b_3}\,,\quad 
\g=-\frac{4[(\tilde a_2+3\tilde a_3)-c^2(\tilde b_2+3\tilde b_3)]}{c^2(\tilde b_1+4\tilde b_2+6\tilde b_3)}\,,\quad 
\l_2=\frac{2(\g-2)}{\g}(\a_2 + 3\a_3)\,, 
\ee
where $\tilde a_n=\o^{-2n}a_n$, $\tilde b_n=\o^{2n}b_n$ and
$\a_n=(\tilde a_n - c^2 \tilde b_n)(c^2-1)/c^2$.  Notice that one may
trade $\tilde a_1$, $\tilde b_1$ and e.g.\ $\tilde a_2$ for,
respectively, $c^2$, $\g$ and the graviton mass $\l_2$, showing that
these may take any value for this class of potentials. When $\g < 2$,
the $\K_i$ are proportional to the constant asymptotic curvatures of
$g_i$; the explicit expressions are given in
appendix~\ref{sec:solvable}.  Finally, $\omega^2$ is also in general a
free parameter that determines for example $\L_2$ to have $\K_2=0$,
after having fine tuned $\L_1$ to set $\K_1=0$.

The expression (\ref{eq:exterior}) resembles the Schwarzschild-dS(AdS)
solution but with a crucial difference: a $r^\g$ term of magnitude $S$
is present and it  may alter significantly the behavior of the
gravitational field, depending on whether $\g<-1$, $-1<\g<0$ or
$\g>0$.

Before discussing these solutions, let us comment on the Isham-Storey
potential $V_{\text{IS}}$ used traditionally. Since in this case all
$b_n$ vanish, it leads to a singular situation where $c^2\to\infty$
unless an additional fine tuning $\omega^2=2/3$ is performed. Even
choosing this case, the linearized analysis is ill defined due to an
enhanced gauge invariance (see~\cite{us} for the case of $\l_\eta=0$),
and moreover from~(\ref{eq:gengamma}) one has $\g\to\infty$. This is
the reason why only standard Schwarzschild-(A)dS solutions were found.

\medskip

Now, in order to shed light on the physical meaning of the various
constants in the solution let us also compute the total gravitational
energy, as measured with respect to backgrounds 1 or 2.  In the
stationary case this is the Komar energy that can be calculated as a
surface integral on a sphere of large radius $r_{outer}\to\infty$ (see
appendix~\ref{energy}). We find, for the two fields:
\be
\begin{split}
& \E_1 = m_1 + S\,\g\,r_{\text{outer}}^{\g+1}  \\
& \E_2 = m_2 - \frac{c }{\omega^2}\, S \, \g \,  r_{\text{outer}}^{\g+1} \,.
\end{split}
\ee
{}From these expressions we see that only IR modifications with $\g<-1$
will lead to finite total energy.

\paragraph{Case $\g< -1$:}
At very large distance the solution reduce to a maximally symmetric
solution parametrized by $\K_i$. In particular one can set
$\K_1=\K_2=0$ with a single fine tuning, determining the asymptotic
conformal factor $\o^2$ as discussed above, so that the solution
describes asymptotically flat metrics.  Clearly because $\g< -1$, at
large distances gravity is Newtonian, while at short/intermediate
distances, depending on $S$, the presence of the additional spin 2
field has changed the nature of the gravitational force.

Since the large distance behavior is Newtonian, the total energy is
finite; taking $r_{outer}\to\infty$, we find $\E_1 = m_1$, $\E_2 =
m_2$. In black hole solutions like these, $m_1$ and $m_2$ are just
parameters, that can be related to the mass of a material object only
when the solution is considered as the outer part of, for instance, a
star. In the case of standard GR, for a star of radius $R$ and mass
density $\rho$, the total gravitational energy $E$ is the total mass
$M=4\pi R^3 \rho/3$. Here, the interaction with $g_2$ is turned on and
we expect a contribution to this energy given by the interaction term
$Q_1$. Its size should be controlled by $V$ and by the matter itself,
because this interaction energy also is turned on by the source.
Moreover, by dimensional analysis the coefficient $S$ of the $r^\g$
term should also be a function of the \emph{size} of the object, and
not only on its mass.  This can be understood intuitively as the
failure of the Gauss theorem due to the presence of $Q$ in the EOMs
and of the $r^\g$ term in the solution.  Accordingly, the separate
contribution of $Q_1$ to the energy is not expressible as a flux on a
2-surface at infinity, as it happens for the total Komar energy.  The
explicit computation of this interaction energy for a star will be
performed in section~\ref{sec:interior}.

\paragraph{Case $\g> -1$:}
For simplicity, also in this case we set $\K_1 = \K_2=0$ as discussed
above, but note that because the new term induces a curvature $R\sim
r^{\g-2}$, only when $\g < 2$ we have that $g_1$ and $g_2$ are
asymptotically flat and $\omega^2$ can be interpreted as an
asymptotical conformal factor.  For these choices of $\g$, we have a
solution such that $Q_1$ does not vanish rapidly as $r\to\infty$, and
compensates a slow fall-off (or rise!) of the gravitational
field. However on dimensional grounds any fall-off slower than $1/r$
makes the Komar total energy infinite, and indeed when $
r_{\text{outer}} \to \infty$ both $\E_1$ and $\E_2$ diverge making
this configuration physically unfeasible.

If spherically symmetric solutions of infinite energy are surely not
physical, this may only suggest that solutions will not be spherically
symmetric, as it happens in non-abelian gauge theories.  For example
one may speculate that finite energy configurations will arrange in
flux tubes of gravitation at large distance, between sources of type 1
and 2, as suggested by the different signs of $S$ in $J$ and $C$.  In
a similar 'confinement-like' scenario, the term $r^\gamma$ may be
screened dynamically by the self-arrangement of configurations of
matter 1 and 2, so that effectively $S\to0$ at large distances, as
suggested by the full star solutions that we will describe later. 

\medskip

To summarize, we found that exact black-hole solutions are modified in
the IR or in the UV depending on the choice of the potential, and that
this behavior is not captured by the linearized approximation.  There
are even cases where the behavior of the potential is not modified at
all with respect to GR (e.g.\ $\g=2$ or $\g=-1$) while the linearized
approximation still shows a linear term. 

It is also interesting to observe that in the limit in which the
second metric decouples, $M_{pl2}\to\infty$ (i.e.\ $\k\to\infty$), and
assuming that $m_2$ and $S$ remains finite, from the
solution~(\ref{eq:exterior}) we find that the term $S\,r^\g$ remains
in $g_1$ while $g_2$ becomes exactly flat. We will discuss the
decoupling limit as well as the limit $c^2\to1$ in
section~\ref{sec:decoupling} and~\ref{sec:LIlimit}.

\subsection{Comparison with the linearized solution}
\label{sec:vain}

It is interesting to comment on the perturbative origin of the exact
solutions. This can be addressed by looking at the asymptotical
weak-field limit of the solution (for $\g<-1$):
\be
\label{eq:sqrtD}
J,\ K,\,C,\,A\sim \text{const} +O(1/r)\qquad
D\sim O(\sqrt{1/r})\,.
\ee
The crucial observation is that $D$ vanishes more slowly (and
non-analytically) than the other components of the perturbations. As a
result, this solution is not captured by the standard linearization,
where all the perturbations have the same large distance fall-off.

Technically, the origin of this behavior can be traced back to the equation
$(Q_{1,2})^t_r=0$, that is algebraic:
\be
\label{eq:Q12}
D(r)\left[\frac{A(r)C(r)+D(r)^2}{J(r)K(r)}
+\frac{a_1 +4 a_2 r^2B(r)^{-1}+6 a_3 r^4B(r)^{-2}}{b_1 +4 b_2 r^{-2}B(r)+6 b_3 r^{-4}B(r)^2}\right]=0
\ee
This equation can be solved either with $D=0$ (type-II solutions) or
with $D\neq0$ (type-I). In this last case, for the exterior solution,
since it turns out that $B=\o^2r^2$, asymptotically the equation turns
into the definition of the speed of light of $g_2$ (as in
(\ref{eq:gengamma})), while the deviations give the mentioned behavior
of $D\sim1/\sqrt{r}$.

From equation~(\ref{eq:Q12}) we can also understand that standard
linearization (around the LB background) can not distinguish between
type-I and type-II at leading order: considering the standard
perturbative expansion (with parameter $\e$) where $D\sim\e$, this
equation starts from order $\e^2$.  Also, at first order $D$ can be
gauged away: it does not appear neither in the linearized Einstein
tensors (due to separate gauge invariances) nor in the mass terms (due
to $\M_1=0$). At higher orders however one must choose $D=0$,
otherwise there is a constraint on the fields $A$, $C$, $J$, $K$, $B$
that are already determined at previous-orders.  We reach the
conclusion that the standard perturbation theory around the LB
background may only approximate the solutions in the type-II branch
(if any exist: we recall that nontrivial type-II solutions are not
known).

On the other hand, it is interesting that the $r^\g$ term can be
recovered in a semi-linearized approach, where one solves exactly
equation~(\ref{eq:Q12}) and treats the remaining ones perturbatively.
This will be done for the interior star solution and the result
containing the $r^\gamma$ terms can be found in
appendix~\ref{sec:fullint}, e.g.\ equation~\ref{eq:middle}.
Alternatively, if one insists in solving perturbatively all the
equations, the correct result can also be recovered by assuming $D\sim
\sqrt{\e}$ while all the other fluctuations are still of order $\e$,
and retaining the first nonvanishing order.\footnote{A similar
  approach was envisaged in~\cite{VAIN} to find an asymptotically flat
  modified Schwarzschild solution for LI massive gravity, valid in the
  $m^2\to0$ limit.  Also in that case a field that is not determined
  at linearized level for $m=0$, is found to vanish non-analitically
  as $\sqrt{1/r}$.  However that solution is not valid beyond some
  distance scale, and there is probably no global
  extension~\cite{VAIN}. In the present work, it is remarkable that
  the semi-linearized solution is also extendable to the exact one.}

Exactly as in the comparison between the Newtonian and Schwarzschild
solutions, the final difference between this semi-linearized and the
exact solution is just that $K=1+2\Phi$ instead of
$K=J^{-1}=1/(1-2\Phi)$ (and similarly for $g_2$).

\subsection{Interior solution}
\label{sec:interior}

In order to determine the integration constants $m_1, m_2$ and $S$ one
can imagine that (\ref{eq:exterior}) is the exterior portion of the
solution describing a spherically symmetric star. We aim at finding
the interior solution and then determine $m_1$, $m_2$, $S$ by matching
with the exterior one.

It is instructive to consider first in full generality a spherical
star made of fluids of type 1 and 2, extending from the origin to
radii $R_1$, $R_2$, stationary with respect to the respective
metrics:
\be
T_1{}_\m^\n=\PM{-\rho_1\\&p_1\\&&p_1\\&&&p_1}\,,\qquad
T_2{}_\m^\n=\PM{-\rho_2&\frac DC (p_2+\rho_2)\\0&p_2\\&&p_2\ \ \\&&&\ \ p_2\ }\,.
\ee

Like in the vacuum, since $g_1$, $T_1$ and $E_1$ are diagonal, so
should be $Q_{1/2}$, i.e.\ $Q_1{}^t_r=0$.  This equation, being the
same as in the vacuum case, is exactly solvable for the class of
potentials~(\ref{eq:genpot}).  The remaining equations are more
involved in the presence of matter, but in linearized approximation
the solution can be found analytically. 

According to the discussion of section~\ref{sec:vain} this partial
linearization corresponds to choosing the type-I class ($D\neq0$) also
for the interior solution.


For simplicity, we consider a star made of an incompressible fluid of
constant density and small pressure, $p\ll\rho$.  The interior
solution is then matched by requiring continuity of $C$, $J$, $B$, $K$
and of the derivatives $C'$, $B'$. This procedure, in the physical
case when $\g<-1$, determines exactly the exterior constants $m_1$,
$m_2$, $S$.\footnote{In the unphysical case $\g>-1$, although the
  potentials $J$ and $C$ are regular, the field $B$ develops a
  singularity $r^{\g+1}$ in the origin, calling probably for a fully
  nonlinear interior solution.}

While the detailed solution is given in appendix~\ref{sec:fullint}, we
present here the instructive case $R_1=R_2=R$, $M_1$,$M_2\neq0$, and
then discuss the phenomenologically interesting case of only matter 1,
$M_2=0$.

\paragraph{Case with both kinds of matter:} For $R_1=R_2=R$, and setting
$M_{1,2}=4\pi\rho_{1,2} R_{1,2}^3/3$, the matching condition gives:
\bea
\qquad 
m_1&=& M_1 + \Delta M\,,\qquad\qquad \qquad \Delta M = \a\,\mu^2
R^2\left(M_1-\frac{\o^2}{\k}M_2\right)\nonumber \\
\label{eq:matchM1M2}
m_2&=& M_2 - \Delta M/ {c\,\kappa\,\o^2}\\[1.3ex]
S  &=& \quad \Delta M\, R^{-(\g+1)}\;\; 15/(2\g-1)(\g-4)\,,\nonumber 
\ena
where $\a=8c^{1/2}\o^2/5(\g+1)(\g-2)$ and $\m^2$ is the same constant
that appears in the linearly growing potential (\ref{eq:muterm}) of
the linearized analysis.  In the exterior solution (\ref{eq:exterior})
$S$ and $\Delta M$ modify the form of standard Schwarzschild solution.
The first modification is in the Newtonian terms, and amounts to a
\emph{mass shift} with respect to the standard values
$m_{1,2}=M_{1,2}$. The second is the \emph{new term} $r^\gamma$.  Both
are proportional to the same combination $\Delta M$.\footnote{But see
  appendix~\ref{sec:fullint} for the full case $R_1\neq R_2$.}

The mass shift $\Delta M$ can be understood as the contribution of the
interaction terms $Q$ to the total energy, i.e.\ to the total mass as
measured by the Newton law at large distance.  To clarify this, it is
useful to recall that in standard GR the Komar energy, written as a
spatial volume integral $(8\pi)^{-1}\int R\xi{\rm d}v$ (see
appendix~\ref{energy}) can be rewritten as a volume integral of the
matter energy-momentum tensor by means of the Einstein equations:
$\E=\E_T=(8\pi)^{-1}\int (2T_\m^\n-T\delta_\m^\n)\xi^\m {\rm d}v_\n$.
This result is modified in massive gravity because the Einstein
equations contain the additional energy-momentum tensor of
interactions $Q$, and the additional contribution can be evaluated
with its volume integral $\E_Q=(8\pi)^{-1}\int
(2Q_\m^\n-Q\delta_\m^\n)\xi^\m {\rm d}v_\n$.  We remark that while the
sum of these two integrals, being the total energy, can be expressed
as a surface integral at infinity (see appendix~\ref{energy}), they
separately can not, and they can only be evaluated using the smooth
interior and exterior solution. The result of the volume integral is
(again finite only for $\g<-1$): \bea \E_{T_{1,2}}= &=&M_{1,2}\,,
\qquad \E_{Q_1}=-\k \,\E_{Q_2} =\Delta M \,.  \ena This confirms that
the mass shift $\Delta M$ is a screening effect, due to the energy of
the interacting fields in $Q$, and corresponding to the nonzero Ricci
curvature even outside the source~(see e.g.~\cite{gabadadze}).

As a side remark, looking at the matching (\ref{eq:matchM1M2}), we
observe that we did \emph{not} linearize in $V$, but neglecting terms
higher order in the matter density one has effectively neglected
higher orders in $V\sim\m^2$. Indeed, the result depends on the two
dimensionless parameters $R^2\m^2$ and $GM/R$, but at first order in
$GM/R$ only the first order in $R^2\m^2$ appears, i.e.\ $GMR\m^2$, and
the final result is smooth when the interaction vanishes,
$V\sim\m^2\to0$. This is opposed to the singular massless limit of
Lorentz-Invariant (Fierz-Pauli) massive gravity.

\paragraph{Case of normal matter:} Turning off  $M_2=0$, we can focus
on sector 1 and discuss more phenomenologically how normal gravity is
modified by the presence of the additional spin two field.

{}From the matching condition we have, with $M_1=M$:
\be
\label{eq:matchM1}
\begin{split}
m_1=&\,  M(1+\a\,\mu^2 R^2)\,, \qquad \qquad m_2=  -\a\,\mu^2 R^2 M/ {c\,\kappa\,\o^2} \\[1ex]
S  =&\,  \m^2 M R^{1-\g}\,\;\; 15\a/(2\g-1)(\g-4)
\end{split}
\ee
and from~(\ref{eq:exterior}) we find for the modified potential (ignoring the numerical factors):
\be
\Phi \sim  G\,M\,\left[\frac1r(1+\m^2R^2) + \m^2 R\left(\frac rR\right)^\g\right]\,.
\ee
The mass shift is now equivalent to a rescaling of the Newton constant
$G(1+\a\m^2R^2)$, that depends on the source radius!!  

We observe that for the sun\footnote{$R_\odot\simeq 5\cdot
10^5\,$Km$\,=5\cdot 10^{15}\,$eV$^{-1}$ and $M_\odot=10^{66}\,$eV.}
we have $\m^2R^2\sim10^{-10}$, assuming all coupling constants to be
of the same order so that $\m\sim m_g\lesssim (10^{-20}\,$eV$)\sim
(100$AU$)^{-1}$ (this limit corresponding to the rough experimental
bound on the graviton mass from pulsar GW
emission~\cite{DUB,DUBA,us}). We thus see that for the sun the size
dependence is negligible and unobservable, and even more so for the
planets. The effect becomes important for objects of size
$\m^{-1}$.  For instance, for large objects with $R\gtrsim10^5 R_{sun}$
(red giants, large gas clouds, galaxies\ldots) the effect may be of
order one, and induces a macroscopic modification of the Newton
constant. For low density objects that we consider here, this
modification does not depend on the mass but just on the object size;
therefore, given the mass, a large sphere of gas has a larger
effective newton constant. In the limit $\m^2R^2>1$, the surface
potential would even scale as $R^4$, instead of $R^2$ as in standard
gravity.  Moreover, remembering that $\m^2$ may be negative, the
negative interaction energy could cause large fluids to antigravitate,
hinting toward the acceleration of the cosmological solutions.

Then, the new term in the potential is of the form
\be 
\delta\Phi \sim G\,M\,\m^2R\left(\frac rR\right)^\g\,, 
\ee
replacing the linear term $G\,M\,\m^2 r$ of the linearized analysis.
The Newtonian and the new term will be competing at a critical
distance $r_c$ that also depends on $\g$:
\be
r_c=R\left|\frac{\m^2R^2}{1+\m^2R^2}\right|^{-\frac1{\g+1}}\,.
\ee
Of course since $\g<-1$ the relevant modification is ultraviolet, and
is evident for $r<r_c$ (while for $\g>-1$ it would be infrared, for
$r>r_c$).  To estimate $r_c$, we observe that since the exponent
$-1/(\g+1)$ is positive, one always has $r_c<R$ for $\m^2>0$, and the
critical distance is inside the star.  This does not mean that there
will be no observable effects, since even subleading modifications to
the newton potential may be measured (for example modifications of the
gravitational potential of relative magnitude $10^{-3/-5}$ are at the
level the current solar-system tests).  On the other hand for negative
$\m^2$, and in particular for $\m^2R^2<-1/2$, one has $r_c>R$ so that
in a UV region near the source the gravitational potential has
stronger fall-off.  For $\m^2R^2\simeq-1$ we even find that $r_c$
becomes infinite, so that the region of UV modification expands to
larger and larger distances!

\pagebreak[3]

We can summarize the results in the physical phase $\g<-1$:
\begin{itemize}
\item For sources of dimension $R<\m^{-1}$, the effects are: a mass
  shift equivalent to a small Newton-constant renormalization
  $(1+\m^2R^2)$, and a subleading correction to the Newtonian
  potential, $\delta\Phi\lesssim(r/R)^\g$.
\item For large sources, of dimension $R>\m^{-1}$, the mass shift is
  more pronounced, and for negative $\m^2$ even the new $r^\g$ term
  can become dominant in a region near the source.
\end{itemize}
As we see, even discarding the nonphysical and possibly confining
branch $\g>-1$, the phenomenology of these modified static solutions
appears to be quite rich, and deserves a separate analysis to confront
their features with real physical systems, e.g.\ modified galactic
gravitational field, gravitation of large sources, post-Newtonian
analysis.

\subsection{Decoupling the second metric}
\label{sec:decoupling}

The idea of introducing a second metric and considering its decoupling
limit, to have a second background at hand while disposing of its
fluctuations, is not new and was indeed considered to tackle the
problem of the nonlinear continuation of the FP massive
gravity~\cite{DAM}. However, due to the singular Isham-Storey
potential, or due to the ill-defined nature of the Lorentz-Invariant
theory, this did not lead to significant advance.  In this work we
found some nonperturbative solutions of the full system, so we are in
a position to control the decoupling limit $M_{pl2}\to\infty$
($\k\to\infty$) in which the second gravity is effectively
switched~off.

First, as far as the propagating states are concerned, we recall from
the linearized analysis~\cite{us} that in the flat background out of
two gravitons only the first graviton survives the decoupling limit:
it is massive (with two polarization states and mass $G\l_2$) and has
a normal dispersion relation.

For the nontrivial solutions, as anticipated, one may take this limit
in the exterior solutions, once one checks that $m_1$, $m_2$ and $S$
stay finite. This is indeed shown by the interior
solution~(\ref{eq:matchM1M2}), therefore we directly find the result:
\bea
 \Phi_1&=&\frac{GM_1}{r}(1+R^2\mu^2\a) + GM_1\mu^2R\left(\frac {r}{R}\right)^\g \left[15\a/(2\g-1)(\g-4)\right] \,,\\
 \Phi_2&=&0\,.
\ena
Both the mass shift and the new term remain, but the second gravity
disappeared: here the other metric is flat!


A look at the exact solution~(\ref{eq:exterior}) in the decoupling
limit shows that the limiting metric 2 is still nondiagonal
($D\neq0$).  This means that $g_2$ is only \emph{gauge-equivalent} to
$\eta_2=\o^2\diag\{-c^2,1,1,1\}$, and that to make contact with this
traditional minkowski diagonal vacuum one has to choose the
gauge~(\ref{eq:g2schw}), where $\bar g_1$ is not diagonal. Explicitly
we find:
\be
\label{eq:g1dec}
 ds_1^2=-\bar J\, dt^2+2\bar D\, dt\,dr+\bar K\, dr^2+r^2d\Omega^2\,,\qquad 
 ds_2^2=\o^2(-c^2 dt^2+ dr^2+r^2d\Omega^2)\,,
\ee
with $\bar J=J$, $\bar K=J^{-1}(1-\bar D^2)$, $\bar D=-(c\o)^{-1}J\sqrt{\o^2-A}$.
Notice that $\bar D$ is defined by the deviation of $A$ from $\o^2$, and
that still $\bar J\bar K+\bar D^2=1$.  

We therefore note that, to recover the present solutions in effective
massive gravity theories, where only $g_1$ is dynamical and the
Lorentz breaking is an external diagonal metric, one should look for
nondiagonal configurations.

We also remark that while taking the decoupling limit has left us with
a flat auxiliary metric, still there is curvature for metric 1 in the
vacuum outside the sources, due to $Q_1$ and $Q_2$ being nonzero
there, because of the $r^\g$ term.  Therefore the order of the limit
matters, and one would not get the correct result if one were to
assume a flat second metric \emph{before} taking the decoupling limit.
In other words, setting $g_2$ to be flat in advance: $g_2= \bar{g}_2$,
we have in vacuum
\bea
&&E_{2\nu}^{\mu}=0= V\delta_\nu^\mu+4(V'X)_\nu^{\mu}\quad\text{so that}\\[.7ex]
&&M_{pl1}^2 E_{1\nu}^{\mu}=2\;V\delta_\nu^\mu \qquad \longrightarrow \nonumber V=\text{const by Bianchi}\longrightarrow
\text{(Anti)deSitter}\,.
\ena
Instead, in the limit $M_{pl2}\to\infty$ we have still
$g_{2\mu\nu}\to\bar{g}_{2\mu\nu}$, but different solutions for $g_1$:
\bea
&&E_{2\nu}^{\mu}\to0  \qquad \text{but}\quad  V\delta_\nu^\mu+4(V'X)^\m_\n\neq 0\,, \qquad\text{and then}\\[.7ex]
&& M_{pl1}^2 E_{2\nu}^{\mu}=(V\delta_\nu^\mu-4V_\nu^{'\mu})\neq
\text{const}\qquad \longrightarrow\text{non-trivial solutions.}\nonumber
\ena

Summarizing, the decoupling limit shows that the theory remains well
behaved, consisting of a modified gravity with massive gravitons,
while the auxiliary metric is flat and decoupled.

\subsection{Lorentz-Invariant limit}
\label{sec:LIlimit}

The Lorentz Invariant limit $c^2\to1$ is also interesting to address
the Vainshtein's claim that nonlinear corrections actually cure the
discontinuity problem in Pauli-Fierz theory~\cite{VAIN}.  Indeed, we
find that the limit $c^2\to1$ is well behaved, and the solutions
retain their validity.  In this limiting phase therefore, gravity is
modified, but lorentz breaking disappears.

The linearized mass term is accordingly of the form $a\,
h_{\m\n}^2+b\,h^2$, however the limiting theory reached in this way is
\emph{not} the Fierz-Pauli one, where $a+b=0$ (and for this reason FP
is free from coupled ghosts). Here, we get to a theory where $a=0$; in
fact, since we approach the LI phase from the $\l_1=0$ branch, and
because in the LI limit one has $\l_1=\l_2=a$, we see that also the
graviton mass vanishes in this limit, as can be checked with the
expression~(\ref{eq:gengamma}). It is nevertheless worth to point out
that also $a=0$ is a ghost-free theory like $a+b=0$. This case is not
usually considered because at the linearized level there is no massive
graviton as a consequence of an additional gauge symmetry (three
transverse diffeomorphisms), see~\cite{alvarez} and also the PF0 phase
in~\cite{us}. Accordingly, no strong coupling problems are expected
and no Vainshtein issues. This matches nicely with our model having
good properties along all the LB branch, that survive also in the LI
limit.

\subsection{Local Lorentz-breaking in nontrivial background}
\label{sec:localLB}

While the asymptotic biflat metrics are Lorentz breaking, one may ask
about the situation at finite distance.  This will have definite
interest when addressing the nonpropagation of ghosts in the described
nontrivial background. In fact, we recall (section~\ref{sec:lin}) that
on flat background this is a consequence of $\M_1=0$ , and this
follows from gauge invariance together with the fact that locally
boosts are spontaneously broken. Now, even in nontrivial background a
spontaneous breaking of Lorentz will lead to flat directions of the
potential.  Whether this fact will be enough to lead to absence of
ghosts and to stable configurations is under scrutiny, and goes beyond
the scope of the present work.

To describe the local breaking of Lorentz at any given point in the
nontrivial background, one chooses a local Lorentz frame ($g_1=\eta$)
and simultaneously diagonalizes $g_2$.  The Lorentz breaking is given
by the entries of $g_2$, that are actually the eigenvalues
of~$X$. These are easily calculated (in polar coordinates $t$, $r$,
$\theta$, $\phi$):
\bea
&\hat X=\o^2\big\{c^2\xi^{-1},\xi,1,1\big\}\,,\\[1.5ex]
&\text{with}\ \xi=\frac12\left[c^2f_-+f_++\sqrt{(c^2-1)(c^2f_-^2-f_+^2)}\right],
\qquad f_\pm=1\pm\tilde Sr^{\gamma-2}\,.\nonumber
\ena
Quite remarkably, that they do not depend on the masses $m_{1,2}$ of
the newtonian terms, and this is due to the nondiagonal structure
given by $D$.  One can easily check that for $r\to\infty$ we have
$\hat X=\o^2\{c^2,1,1,1\}$, reproducing the asymptotical lorentz
breaking ($\g<-1$).

Then we see that in the case $S=0$ the eigenvalues are constant, so
that at any distance the Lorentz breaking is the same: $\hat
X=\o^2\{c^2,1,1,1\}$. We have two pure Schwarzschild solutions in a
configuration that at any point breaks local boosts but preserves
rotations (these are the solutions found in~\cite{alvarez}).

On the other hand, since in general $S\neq0$, a star solution will
break not only boosts but also local rotations, because the $rr$ term
is different from the $\theta\theta$ and $\phi\phi$ ones. For example
at large but finite distance, where $S r^{\gamma-2}$ is small, we have:
\be
\hat X\simeq\o^2\{c^2(1-\tilde S r^{\gamma-2}),1+\tilde S r^{\gamma-2},1,1\}\,.
\ee

To compare the situation with standard GR, we recall that in GR
Lorentz-invariance at any given point is always valid, in the Lorentz
frame, and is broken in a finite neighbourhood only by the curvature
(tidal) effects.  Here on the contrary in the gravitational sector a
Lorentz breaking is felt also locally, and for $\tilde S\neq 0$ also
rotations are broken. The physical effect is that gravitons will propagate
differently in direction of the source.

We strongly believe that this breaking of (local) boosts \emph{and}
rotations at finite distance from a source is a general feature of
nontrivial solutions in massive gravity, due to the presence of
additional fields that can not be `gauged away'.

\pagebreak[3]

\section{Conclusions}

In this paper we approached the problem of finding a consistent
massive deformation of gravity by introducing an additional spin 2
field $g_2$ coupling non-derivatively to the standard metric field.
This allows us to explore both the Lorentz invariant (LI) and Lorentz
breaking (LB) phases working with consistent and dynamically
determined backgrounds.  Preserving diffeomorphisms and breaking
Lorentz is also important; at the linearized level it forces $\M_1$ to
vanish and no dangerous scalar mode is propagating.  Still at the
linearized level, it was shown~\cite{DUB} that in the case $\M_1=0$
the vDVZ discontinuity is absent, but a new linearly growing term is
present in the static gravitational potential~\cite{us,massivegr},
that seems to invalidate perturbation theory beyond some distance
scale.  To address this and the vDVZ discontinuity problem, we thus
studied the exact spherically symmetric configurations.  The exact
solution that we found, valid for a large class of interaction
potentials, shows that the linear term is replaced in the full
solution by a power-like term $r^\gamma$, with $\g$ depending on the
nonlinear couplings in the interaction potential.

Phenomenologically, when $\gamma <-1$ the total energy of the solution
is finite and the space is asymptotically flat; Lorentz is broken in
the gravitational sector by the asymptotic value of $g_2$, but normal
matter only feels the modification of the gravitational potential.
Using the full solution one can check that the absence of the vDVZ
discontinuity is an exact result. The effect of the interaction
manifests in the $r^\gamma$ term whose size $S$ was determined for a
star by matching the exterior solution with an interior one.  In
addition to this, by the presence of the additional spin 2 field, the
total mass of the star appearing in the Newton term gets a finite
renormalization that depends on the object size, and may screen or
even antiscreen the star mass.  We believe that this is a general
feature of massive gravity.

When $g_1$ describes a black hole the solution depends not only on the
collapsed mass but also on an other constant, probably remnant of the
original shape; notice that there is no contradiction with the no-hair
theorem because the Einstein equations are modified by the presence of
$Q_{1/2}$.

In the case $\gamma >-1$ the total energy is infinite and this may
indicate only that solutions will not be spherically symmetric. For
example the solution may be unstable under axially symmetric
perturbations and drop to a flux tube in a sort of mass confinement
scenario.

Indeed, regarding stability, even for the physical case $\g<-1$ the
final word would be given by studying the small fluctuations also
around the exact solution, to check that the non-propagation of the
(ghost) scalars and vectors is preserved on a nontrivial
background. To this aim, we have discussed how the spontaneous
Lorentz-breaking is present also in the nontrivial background, where
we note that in general also rotations are locally broken. One expects
this also to be a generic feature of massive gravity.

We showed that we can reach the LI phase by tuning $c^2\to1$ in the
exact solutions, and this results into a well behaved phase, though
not the Fierz-Pauli one (gravitons are massless). The fate of the
discontinuity and Vainshtein claim for the PF case is thus still an
open problem and exact solutions of type II with $c=1$ are presently
under investigation.  Finally, it would be interesting to speculate on
the role of the mass screening in cosmology, that may change the form
of the Hubble expansion.

\acknowledgments 

Work supported in part by the MIUR grant for the Projects of National
Interest PRIN 2006 ``Astroparticle Physics'', and in part by the
European FP6 Network ``UniverseNet" MRTN-CT-2006-035863.

\begin{appendix}

\section{Background for solvable potentials}
\label{sec:solvable}

For the solvable potential~(\ref{eq:genpot}), we report here the
biflat solution as it results from solving the EOM~(\ref{bc}), as well
as the $\m^2$ and $\l_2$ constants of the linearized analysis.

The fine tuning conditions to ensure flatness $\K_1=\K_2=0$ turn in
two relations for the two cosmological constants. Defining $\tilde
a_n=\o^{-2n}a_n$, $\tilde b_n=\o^{2n}b_n$ and $\a_n=(\tilde a_n - c^2
\tilde b_n)(c^2-1)/c^2$, and $\b_n=(\tilde a_n + c^2 \tilde b_n)$, we have:
\bea
3 \K_1 &=&{} -2\o^{-2}  \L_1 +\frac{c^{-3/2}}{  (c^2-1) \g}
   \bigg[
    -8 c^2 (3 c^2+\g +1) \a _2
    -12 c^2 ((\g +6) c^2+3 \g +2) \a _3\bigg]\nonumber \\
 && {}\qquad\qquad+ c^{-3/2}\bigg[(6 c^2-2) \beta _2+24
   (c^2-1) \beta _3-\tilde{a}_0 c^2+3\tilde{b}_4 c^4-5 \tilde{a}_4\bigg]
\nonumber\\
3 \K_2&=&{}- 2  \o^2\L_2  \k^{-1} +\frac{c^{-5/2}}{
   (c^2-1) \g  \k }\bigg[ 8 ((\g +1) c^2+3) \a _2 c^2+12
   ((3 \g +2) c^2+\g +6) \a _3
   c^2\bigg]\nonumber \\
&&{}\qquad\qquad - c^{-5/2}\bigg[2 (c^2-3) \beta _2+24
   (c^2-1) \beta _3+\tilde{a}_0 c^2+5 \tilde{b}_4 c^4-3 \tilde{a}_4)\bigg]
\label{eq:flatness}
\ena
and we remind that one of these is a genuine fine tuning to achieve
flatness, as is usual in General Relativity, while the other is a
complicated equation that may be used to find $\o$. We prefer thinking
in reverse and consider $\o$ a free parameter determining the right $\L_2$.
Finally, the lorentz breaking speed of light turns out to be
\be
 c^2=-\frac{\tilde a_1+4\tilde a_2+6\tilde a_3}
           {\tilde b_1+4\tilde b_2+6\tilde b_3}\,.
\ee

The relevant quantities entering in the linearized analysis are the
graviton mass $\l_2$ and the $\m^2$ parameter, that have the following
expressions: 
\bea
M_{pl1}^2\mu^2 &=& \frac{(\g-2)^2}{32} \bigg \{  \tilde{a}_0 -\frac3{c^{2} (c^2-1)(\g-2)\g}\left[ 5 (c^2-1)(\g-2)
  \g(\tilde{a}_4 + c^4 \tilde{b}_4)  \right. \nonumber\\
 &&\qquad\qquad\quad\left. \left.{} + 6 c^2 (1 + c^2)(\g-2) \g \beta_2 + 32 c^2(1 + c^2)(\g-2)\g
 \beta_3  \right. \right.\nonumber\\
&&\qquad\qquad \quad\left. {}-4 c^2 (c^2-1)\left[ (\g+2)^2 \alpha_2 + (12 + 4\g
    + 7 \g^2)\alpha_3 \right] \right]\bigg\} \\
M_{pl1}^2m_g^2=\l_2&=&G\frac{2(\g-2)}{\g}(\a_2 + 3\a_3)\,.
\ena

\section{Interior solution}
\label{sec:fullint}

For generality we report here the case of a star composed of two
spherical regions filled with incompressible fluids of kind 1 and 2
extending from the origin to different radii $R_1$ and $R_2$; of constant
densities $\rho_{1,2}= \frac{M_{1,2}}{4/3\, \pi R_{1,2}^3}$ and
negligible pressures.  In general we have two scenarios, for the three
different regions:
\smallskip

$a$) for $\;R_2<R_1$ we have $0<r<R_2$, $R_2<r<R_1$ or $r>R_1$;

$b$) for $\;R_1<R_2$ we have $0<r<R_1$, $R_1<r<R_2$ or $r>R_2$.

\smallskip
 
\noindent
We give only the analitic results for $J[r]$, that is the gravitational
potential in $g_1$:

\begin{itemize}
\item Starting from the exterior solutions $r>R_{1,2}$, we find a
common value for 
the exterior potential $J$ in both
scenarios $a$) and $b$):
\bea\nonumber
J(r)=&1&{}- \frac{2 G }{r}\left[M_1+\frac{16 \mu ^2  \left(M_1 R_1^2-\omega ^2 \k^{-1}M_2 R_2^2\right)\sqrt{c}\,\omega ^2}{5 
   (\gamma -2) (\gamma +1) }\right]+\\
&&{}+ r^{\gamma }\left[\frac{96 G \mu ^2  \left(\omega ^2 M_2 R_2^{1-\gamma }-\kappa 
   M_1 R_1^{1-\gamma }\right)\sqrt{c}\,\omega ^2}{(\gamma -4) (\gamma -2) (\gamma +1) (2 \gamma -1) \kappa }\right] \,.
\ena

\item
The intermediate solutions are different in the two scenarios:

for $a$) i.e.\ $R_2<r<R_1$
\bea
1&- &\frac{3 G M_1}{R_1}+\frac{32 G \sqrt{c} \mu ^2 \omega ^4 M_2 R_2^2}{5 r (2-\gamma ) (\gamma +1) \kappa }+\frac{G r^2 M_1 \left(1-\frac{16 \sqrt{c} \mu ^2 \omega ^2 R_1^2}{(2-\gamma ) (\gamma
   +1)}\right)}{R_1^3}-\frac{48 G r^4 \sqrt{c} \mu ^2 \omega ^2 M_1}{5 (\gamma -4) (\gamma
   +3) R_1^3}+\nonumber\\
&&\frac{96 G \sqrt{c} \mu ^2 \omega ^2 M_1 R_1^{\gamma } r^{1-\gamma }}{(\gamma -2) (\gamma +1)
   (\gamma +3) (2 \gamma -1)}+\frac{96 G \sqrt{c} \mu ^2 \omega ^4 M_2 R_2^{1-\gamma }
   r^{\gamma }}{(\gamma -4) (\gamma -2) (\gamma +1) (2 \gamma -1) \kappa }
\label{eq:middle}
\ena
for $b$) i.e.\ $R_1<r<R_2$
\bea\nonumber
1&-&\frac{2 G M_1}{r}-\frac{32 G \sqrt{c} \mu ^2 \omega ^2 M_1 R_1^2}{5 r (2-\gamma ) (\gamma +1)}-\frac{16 G r^2 \sqrt{c} \mu ^2 \omega ^4 M_2}{(\gamma -2) (\gamma +1) \kappa  R_2}+\frac{48 G r^4 \sqrt{c} \mu ^2 \omega ^4 M_2}{5 (\gamma -4) (\gamma +3) \kappa  R_2^3}-\\
&&\frac{96 G \sqrt{c} \mu ^2 \omega ^4 M_2 R_2^{\gamma } r^{1-\gamma }}{(\gamma -2) (\gamma
   +1) (\gamma +3) (2 \gamma -1) \kappa }-\frac{96 G \sqrt{c} \mu ^2 \omega ^2 M_1
   R_1^{1-\gamma } r^{\gamma }}{(\gamma -4) (\gamma -2) (\gamma +1) (2 \gamma -1)}.
\ena

\item The inner solutions $r<R_{1,2}$ have again a common form in both scenarios $a$) and $b$):
\bea\nonumber
1-\frac{3 G M_1}{R_1}+G r^2 \left[M_1 \left(\frac{16 \sqrt{c} \mu ^2 \omega ^2}{(\gamma -2) (\gamma +1)
   R_1}+\frac{1}{R_1^3}\right)-\frac{16 \sqrt{c} \mu ^2 \omega ^4 M_2}{(\gamma -2) (\gamma
   +1) \kappa  R_2}\right]+\\
\frac{48 G \sqrt{c} \mu ^2 \omega ^2 \left(\omega ^2 M_2 R_1^3-\kappa  M_1 R_2^3\right)
   r^4}{5 (\gamma -4) (\gamma +3) \kappa  R_1^3 R_2^3}
+\frac{96 G \sqrt{c} \mu ^2 \omega ^2 \left(\kappa  M_1 R_1^{\gamma }-\omega ^2 M_2
   R_2^{\gamma }\right) r^{1-\gamma }}{(\gamma -2) (\gamma +1) (\gamma +3) (2 \gamma -1) \kappa
   }.\ \ \ 
\ena
As one checks, the solution is regular at the origin.
\end{itemize}

\section{Energy integrals}
\label{energy}

In the presence of a time-like Killing vector in GR on can define the
notion of total gravitational energy as a flux from an asymptotic
2-surface that involve only the gravitational field at large distance,
far from the sources.  Consider the following metric
\be
ds^2 = -C(r) \, dt^2 + 2 D(r) \, dr dt  + A(r) \, dr^2 + B(r) \, d \Omega^2 \,,
\ee
with the time-like Killing vector $K = \frac{\de}{\de t}$. From the
Killing equation we have
\be
\nabla^\mu J_\mu = 0 \, , \qquad J_\mu = \Box K_\mu \, .
\ee
The Komar energy $\E$ is defined by
\be
\E = w \int_{t = t_1} \sqrt{h}  \,  n^\mu J_\mu  \,,
\ee
where $h_{\mu \nu}$ is the induced metric in the hyper-surface $t=
const.$ with unit normal $n^\mu$ and $w$ is a normalization
constant. According to Stokes theorem, given a 3-surface $V$, for any
antisymmetric tensor $F_{\mu \nu} $ we have
\be
\int_V  d^3 x \, \sqrt{h} \, n^\mu \nabla^\nu F_{\mu \nu} = 
\int_{\partial V}  d^2 x \, \sqrt{\g}  
\,  \left( n^\alpha  v^\beta - n^\beta v^\alpha   \right) \, F_{\alpha \beta} \,,
\ee
where $v^\alpha$ is the unit normal to $\de V$ and $\g_{\alpha
  \beta}$ is the induced metric in $\de V$. Then Stokes theorem gives
\be
\E = - \frac{w}{2}
 \, \int_{t = t_1, r = r_1.} \sqrt{\g}  \,  \left( n^\alpha  v^\beta - v^\alpha  n^\beta \right) \nabla_\alpha K_\beta  \,.
\ee     
In general the Komar energy will depend on 2-surface that bounds the
$t=const.$ slice.  Indeed, from Einstein equations it easy to show
that the difference $\Delta \E$ between the Komar energy computed with
two different bounding 2-surfaces $\Sigma_1$ and $\Sigma_2$ is
proportional to the integral of the Ricci tensor over the 3-volume
bounded by $\Sigma_1$ and $\Sigma_2$. As a result the Komar energy
does not depend on $\Sigma$ in a region where the Ricci tensor is
vanishing. This is indeed the case in a region far from any source.
Then
\be
\E = - \frac{w}{2}
 \, \int_{t = const., r \to \infty.} \sqrt{\g}  \,  \left( n^\alpha  v^\beta - v^\alpha  n^\beta \right) \nabla_\alpha K_\beta  \,;
\ee 
 for the induced metric on the 3-surface
$t=\text{const.}$ and its normal $n$ we get 
\be
\begin{split}
& dl^2 =  A(r) \, dr^2 + B(r) \, d \Omega^2  \; ; \\
& n = (AC + D^2)^{-1/2} \left( - A^{1/2} \frac{\de}{\de t} +   D A^{-1/2} \frac{\de}{\de r}
\right) 
\end{split}
\ee
and for the induced metric on $t,r = const.$ and its normal $v$ ( v is
normalized with h )
\be
ds^2 =   B(r) \, d \Omega^2 \, , \qquad \qquad v =   A^{1/2} \frac{\de}{\de r} \;  .
\ee
We have then  
\be
\E = -  \lim_{r\to \infty} \, {\frac{4 \pi w  \, C^\prime \,  B}{\sqrt{D^2 + AC}}}    \; .
\ee  

One can recover the same result using the language of differential
forms.  Introducing the 1-form $J=- (\delta d + d \delta ) \tilde K$
in terms of the 1-form $ \tilde K$ associated with the Killing vector
$K$. From the Killing equation $\delta \tilde K = 0$, then
\be
J = - (\delta d + d \delta ) \tilde K  \equiv \delta d \tilde K = \ast d \ast d  \tilde K
 = \Box K_\mu dx^\mu \; .
\ee
Now
\be
\int d \ast  J = 0 \, \Rightarrow \int_{\text{t=\rlap{const.}}} \ast \tilde J
\ \ \ \text{ is time-independent}
\ee
and finally
\bea
 \E &=& - w \int_{t = t_1} \!\!\!\! \ast J = - w \int_{t = t_1}\!\!\!\! \ast \delta d \tilde K =  w \int_{t = t_1}\!\!\!\! \ast \ast d \ast d \tilde K  =  w \int_{t = t_1} \!\!\!\! d \ast d \tilde K \nonumber\\
 &=& w \int_{t = t_1, r=r_1} \!\!\!\! \ast d \tilde K  = \int_{t = t_1, r=r_1}\!\!\!\!\sqrt{g} \, \frac{C^\prime}{2} \, g^{\mu t} \, g^{\nu r} \, \epsilon_{\mu \nu \alpha \beta} \, \, dx^\alpha \wedge  dx^\beta \nonumber\\
 &=& -   
\frac{4   \pi w \, B C^\prime}{\sqrt{D^2 + AC}}\,.
\ena

\section{Simplest bigravity}
\label{sec:simplest}

Here we analyze the system in the simpler particular case when $V$ is
a function of $q$ only: $V=f(q)$. The Bianchi identities (\ref{cons})
for $Q_{1,2}$, can be written as\footnote{And recall that for any
  vector field $v_\m$ one has $ (\nabla_{2 \mu} - \nabla_{1 \mu})
  v_\nu = C_{\mu \nu}^\sigma v_\sigma$, with the tensor $C_{\mu
    \nu}^\sigma = g_2^{\sigma \beta} \left( \nabla_{1 \mu} g_{2 \nu
    \beta} + \nabla_{1 \nu} g_{2 \mu \beta} - \nabla_{1 \beta} g_{2
    \mu \nu} \right)/2$.}
\bea
\label{eq:action}
 \partial_{\mu}V - \left[ \de_\nu \log q  +
2 \,  \left(\nabla_{1 \nu}-\nabla_{2 \nu} \right) \right]\left(V^\prime X \right)^\nu_\mu = 0 \\[.3cm]
8 \left(\nabla_{1 \nu} + \nabla_{2 \nu} \right)  \left(V^\prime X \right)^\nu_\mu- V \, \partial_{\mu} \log q =0 
\ena
and because in the case at hand $\left(V^\prime X \right)^\nu_\mu =
f^\prime q \delta^\nu_\mu$, the only non-trivial equation is
\be
\left[16 \, \frac{d^2}{d (\log q)^2} f - f \right] \de_\mu q =0 \; .
\ee
Thus, either $q=$const. or $ V =V_0= c_1 q^{1/4} + c_2
q^{-1/4}$. However, $(g_1 g_2)^{1/4} V_0 = c_1 \sqrt{g1} + c_2
\sqrt{g2}$ would imply that the two sectors do not see each other and
we are left with two independent copies of GR + cosmological term. The
theory with $q=$const is the simplest of all possible bigravity
theories, the EOM reduce to
\begin{gather}
M_{pl1}^2 {E_1}^\mu_\nu  +  \K_1 \,  \delta^\mu_\nu 
 = \frac{1}{2} {T_1}^\mu_\nu    \\
M_{pl2}^2 {E_2}^\mu_\nu  +   \K_2  \,  \delta^\mu_\nu =  \frac{1}{2} {T_2}^\mu_\nu   \,,
\end{gather}
 with
 \begin{gather}
 \K_1 = q^{1/4} \left[ f(q)  - 4 q \, f^\prime(q)  \right]  \\
 \K_2 = q^{-1/4} \left[ f(q)  + 4 q \, f^\prime(q)  \right].
 \end{gather}
The effective cosmological constants are thus related; moreover, this
simplest bigravity, due to the constraint $q=$const, is equivalent to
a single GR + unimodular GR, and the two sectors share the conformal
mode.  Finally, besides the diagonal diff also two independent
volume-preserving diffs are present.

\medskip

As an example of exact solution, we present the solution for the potential
\be
V=\tr \ln X=\ln\det X
\ee
(and we may add also the two cosmological constant terms
$\Lambda_1q^{-1/4}$ and $\Lambda_2q^{1/4}$ to achieve flatness).  With
this potential we have $V'X= {\mathbf{1}}$ by construction. The
solution in general is Schwarzschild-deSitter for both metrics, but
$g_2$ is in a different gauge:
\bea 
J&=&\Delta_1\left(1-2\frac{m_1}{r}+ c_1r^2\right)\,,
\qquad\qquad K=\Delta_1/J\,,\\ 
C&=&\Delta_2\left(1-2\frac{m_2}{\rho} +c_2 \rho^2\right)\,, 
\qquad\qquad \rho=(r^3+\lambda^3)^{1/3}\\ 
B&=&\o^2\rho^2 \,,
\qquad D^2+AC=\Delta_2\frac{(B')^2}{B}=c^2\o^4\Delta_1\, (\rho')^2\,,\\ 
A&=&\text{free} \,,\qquad\qquad c^2=\frac{4\Delta_2}{\o^2\Delta_1}\,.
\ena
This is a family of solutions because $A(r)$ is a free function (!),
remnant of the spatial diffs. The determinant $A C+D^2$ is fixed by
$B(r)$, and for $\l\neq0$ it is not constant, at finite distance.
Then one can also use $A$ to set $D=0$ and get a bidiagonal solution
like (\ref{eq:g2schw}).  Notice that $\o^2$ and $\Delta_2/\Delta_1$
are free constants, and so also the relative speed of light $c^2$ is
free.

\end{appendix}

\end{document}